\documentclass[aps,twocolumn,groupedaddress,amsmath,amssymb]{revtex4}

\usepackage{graphicx}
\usepackage{dcolumn}
\usepackage{xcolor}%
\usepackage{booktabs}%
\usepackage{bm}
\usepackage{color, soul}

\begin{document}

\title{Mo$_{3}$ReRuC: A noncentrosymmetric superconductor formed in the MoReRu-Mo$_{2}$C system}

\date{\today}
\author{Qinqing Zhu$^{1,2,3}$}
\author{Guorui Xiao$^{2,3,4}$}
\author{Wuzhang Yang$^{2,3,5}$}
\author{Shijie Song$^{4}$}
\author{Guang-Han Cao$^{4}$}
\author{Zhi Ren$^{2,3}$}
\email{renzhi@westlake.edu.cn}

\affiliation{$^{1}$Ningbo Institute of Technology, Beihang University, 399 Kangda Road, Beilun District, Ningbo, P. R. China}
\affiliation{$^{2}$Department of Physics, School of Science, Westlake University, 18 Shilongshan Road, Hangzhou, 310024, Zhejiang Province, PR China}
\affiliation{$^{3}$Institute of Natural Sciences, Westlake Institute for Advanced Study, 18 Shilongshan Road, Hangzhou, 310024, Zhejiang Province, PR China}
\affiliation{$^{4}$School of Physics, Zhejiang University, Hangzhou 310058, PR China}
\affiliation{$^{5}$Department of Physics, Fudan University, Shanghai 200433, PR China}

\begin{abstract}
A quaternary compound with the composition Mo$_{3}$ReRuC is obtained in a previously unexplored MoReRu-Mo$_{2}$C system.
According to x-ray structural analysis, Mo$_{3}$ReRuC crystallizes in the noncentrosymmetric space group $P$4$_{1}$32 (cubic $\beta$-Mn type structure, $a$ = 6.8107(1) {\AA}). Below 7.7 K, Mo$_{3}$ReRuC becomes a bulk type-II superconductor with an upper critical field close to the Pauli paramagnetic limit. The specific heat data gives a large normalized jump $\Delta$$C_{\rm p}$/$\gamma$$T_{\rm c}$ = 2.3 at $T_{\rm c}$, which points to a strongly coupled superconducting state.
First principles calculations show that its electronic states at the Fermi level are mainly contributed by Mo, Re and Ru atoms and strongly increased by the spin-orbit coupling.
Our finding suggests that the intermediate phase between alloys and carbides may be a good place to look for $\beta$-Mn type noncentrosymmetric superconductors.
\end{abstract}

\maketitle
\maketitle

\section{Introduction}
The inversion symmetry is of importance in determining the band degeneracy away from the high-symmetry points of the Brillouin zone and hence affects profoundly the superconducting pairing.
In centrosymmetric superconductors, the spin degeneracy of Cooper pairs is protected by the presence of inversion symmetry.\cite{anderson1984structure}
In noncentrosymmetric superconductors (NCSs), however, the spin degeneracy of bands is lifted due to the antisymmetric spin-orbit coupling (ASOC) caused by the absence of inversion symmetry.
This can allow for a mixture of spin-singlet and spin-triple Cooper pairing, whose mixing degree depends on the ASOC strength.\cite{bauer2012non,smidman2017superconductivity,gor2001superconducting}
As a consequence, NCSs are predicted to display various peculiar properties \cite{1976,1985,1994,1996,2009,2018}, including magnetoelectric effect, spin-polarized currents, helical states and the Josephson diode effect.
On the experimental side, while these predictions remain to be verified, a wealth of exotic behaviors such as large upper critical fields,\cite{bao2015superconductivity,carnicom2018tarh2b2,gornicka2021nbir2b2} nodal superconducting gap,\cite{bonalde2005evidence,shang2018nodeless} and broken time-reversal symmetry\cite{shang2018nodeless,hillier2009evidence} have indeed been observed.

The cubic $\beta$-Mn type NCSs crystalline in a pair of enantiomorphic space groups, $P$4$_{1}$32 and $P$4$_{3}$32, and hence are not only noncentrosymmetric but also chiral.\cite{lovesey2021structural} These NCSs have attracted much attention since the observation of a spin-triplet pairing component in Li$_{2}$Pt$_{3}$B.\cite{Yuan2006} This contrasts with the case of isostructural Li$_{2}$Pd$_{3}$B, where a spin-singlet pairing is dominant.\cite{Yuan2006}
Later, $\beta$-Mn type Mo$_{3}$Al$_{2}$C is shown to be a strongly coupled superconductor with $T_{\rm c}$ up to 9.3 K and its superconducting behavior exhibits a notable deviation from the BCS-type.\cite{bauer2010unconventional,karki2010structure}
In addition to these two prominent examples, nearly ten $\beta$-Mn type NCSs have been reported so far\cite{toth1966superconducting,khan1973superconducting,kawashima2006superconductivity,iyo2019superconductivity,ying2019superconductivity,zhu2022w} and, except for W$_{4}$IrC,\cite{zhu2022w} they have the general formula $M_{2}$$M'_{3}$$X$ or $M_{3}$$M'_{2}X$.
Here $M$ and $M$' are either both transition metal elements or a transition metal element plus a main group metal element, while $X$ is a nonmetallic light element such as B, C, N, P, and S.
In spite of this progress, however, the exploration of new $\beta$-Mn type NCSs remains challenging since the common strategy requires tests on numerous combinations of the three constituent elements from different parts of the periodic table.\cite{iyo2019superconductivity}

Previously, it has been shown that there exists a continuous solid solution exists Re and Mo$_{2}$C, both of which have a centrosymmetric hexagonal structure.\cite{raub1972superconductivity}
The equiatomic medium-entropy alloy MoReRu\cite{lee2019superconductivity} is analogues to Re in many aspects, including crystal structure, valence electron count and reactivity with carbon.
This striking similarity led us to investigate the MoReRu-Mo$_{2}$C system.
Unexpectedly, a new quaternary compound with the formula Mo$_{3}$ReRuC is identified and found to adopt the noncentrosymmetric $\beta$-Mn type structure.
Physical property measurements show that Mo$_{3}$ReRuC becomes a bulk type-II superconductor below 7.7 K. Various superconducting parameters are obtained and the theoretical band structure is investigated.
These results are compared with those of existing $\beta$-Mn type NCSs and the implications are discussed.

\section{Materials and Methods}
Polycrystalline (MoReRu)$_{1-x}$(Mo$_{2}$C)$_{x}$ samples with $x$ up to 0.6 were prepared by the arc melting method.
Stoichiometric amounts of Mo (99.99\%), Re (99.99\%), Ru (99.99\%) and C (99.99\%) powders were mixed thoroughly and pressed into pellets in an argon-filled glove box. The pellets was then put in an arc furnace and melted several times under high-purity argon atmosphere, followed by rapid cooling on a water-chilled copper plate. The powder x-ray diffraction (XRD) patterns of the samples were collected using a Bruker D8 Advance diffractometer with Cu K$\alpha$ radiation. The refinements of crystal structure were carried out using the JANA2006 program.\cite{petvrivcek2014crystallographic} The sample morphology and composition were investigated in a Hitachi SU-8020 field emission scanning electron microscope (SEM) equipped with an Oxford X-MAX 80 energy dispersive x-ray (EDX) spectrometer. Electrical resistivity and specific heat were measured in a Quantum Design Physical Property Measurement System (PPMS-9 Dynacool). The resistivity measurements were done on bar-shaped samples using a standard four-probe method with an applied current of 1 mA.
The dc magnetization measurements were done in a Quantum Design Magnetic Property Measurement System (MPMS3). First-principles calculations were performed in the Vienna Ab-initio Simulation Package \cite{kresse1996efficient} using the Perdew-Burke-Ernzerhof (PBE)\cite{perdew1996generalized} exchange-correlation functional with the convergence threshold of Hellmann-Feynman force of 0.01 eV/{\AA}. The energy convergence criterion was set to 10$^{-6}$ eV for structural optimization, and was increased to 10$^{-8}$ eV for the other calculations. In addition, the wavefunction cutoff energy was fixed to 450 eV. The $\Gamma$-centered \emph{\textbf{k}} mesh was set to 5$\times$5$\times$5 in both structural optimization and self-consistent calculations, and was increased to 10$\times$10$\times$10 for density of states (DOS) calculations.
\begin{figure}
\includegraphics*[width=8.6cm]{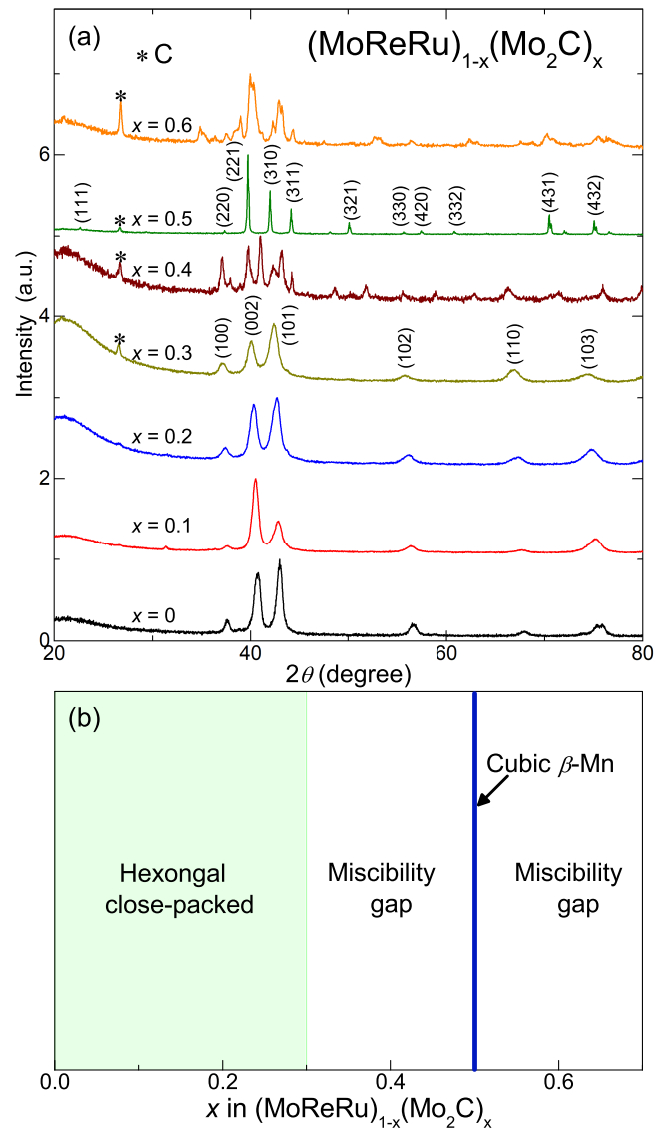}
\caption{
(a) XRD patterns for the series of (MoReRu)$_{1-x}$(Mo$_{2}$C)$_{x}$ samples with increasing $x$ up to 0.6. For $x$ = 0.3 and 0.5, the major diffraction peaks are indexed on the $P$6$_{3}$/$mmc$ (hexagonal) and $P$4$_{1}$32 (cubic) space groups, respectively.
The asterisks mark the impurity peak due to carbon.
(b) Partial phase diagram of the MoReRu-Mo$_{2}$C system at ambient condition.
}
\label{fig1}
\end{figure}

\begin{figure*}
\includegraphics*[width=16cm]{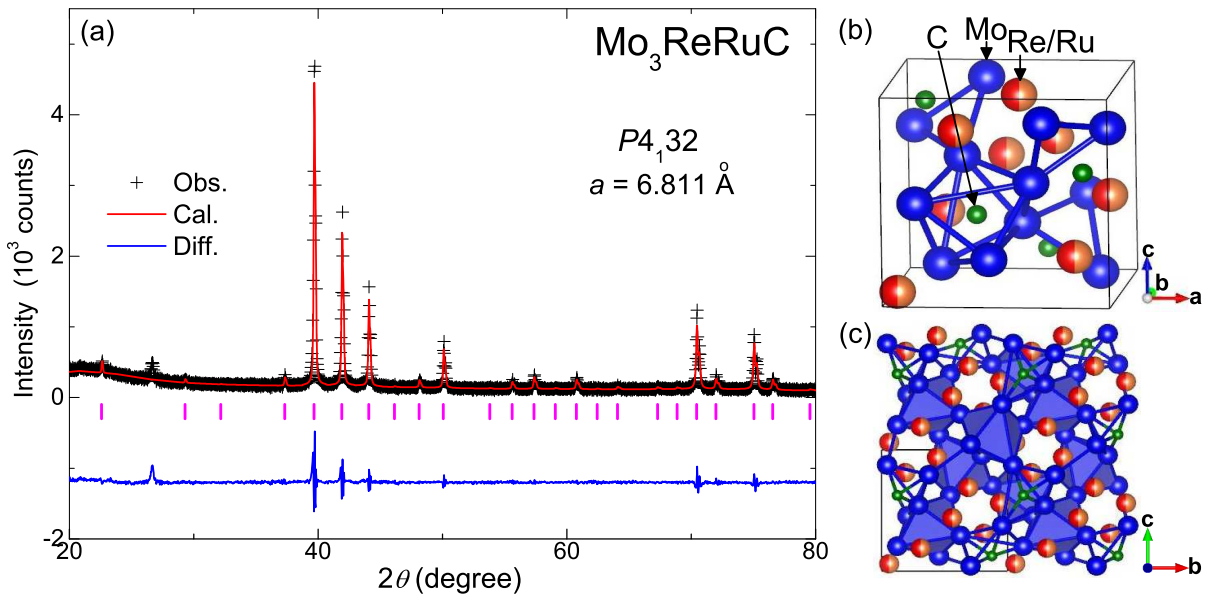}
\caption{
(a) Structural refinement profile for Mo$_{3}$ReRuC.
(b, c) Schematic structure for Mo$_{3}$ReRuC and its projection along the $c$-axis.
}
\label{fig2}
\end{figure*}

\begin{table*}
	\centering
	\caption{Structural refinement results for Mo$_{3}$ReRuC.}
	\renewcommand\arraystretch{1.3}
	\begin{tabular}{p{1cm}<{\centering}p{1.3cm}<{\centering}p{1.8cm}<{\centering}p{1.8cm}<{\centering}p{1.8cm}<{\centering}p{1.8cm}<{\centering}}
		\\
        \hline
        \multicolumn{6}{c}{Mo$_{3}$ReRuC} \\
        \hline
        \multicolumn{3}{c}{Space group} & \multicolumn{3}{c}{$P$4$_{1}$32}\\
        \multicolumn{3}{c}{$a$ ({\AA})} & \multicolumn{3}{c}{6.8107(1)}\\
        \multicolumn{3}{c}{$V$  ({\AA}$^{3}$)} & \multicolumn{3}{c}{315.9(1)}\\
        \multicolumn{3}{c}{$R_{\rm wp}$} & \multicolumn{3}{c}{9.7\%}\\
        \multicolumn{3}{c}{$R_{\rm p}$ } & \multicolumn{3}{c}{7.3\%}\\
        \multicolumn{3}{c}{G.O.F.} & \multicolumn{3}{c}{1.4}\\
		\hline 
		Atoms& site &$x$& $y$ &  $z$ &  Occupancy \\
		\hline
		Mo   & 12$d$ & 0.125 &  0.206 &0.456& 1\\
        Re   & 8$c$ & 0.061 &  0.061 &0.061& 0.5\\
        Ru   & 8$c$ & 0.061 &  0.061 &0.061& 0.5\\
		C & 4$a$ & 0.375& 0.375 &0.375  & 1\\
		\hline 
	\end{tabular}
	\label{Table2}
\end{table*}

\section{Results and Discussion}
\subsection{The MoReRu-Mo$_{2}$C system}
Figure 1a shows the evolution of XRD patterns for the (MoReRu)$_{1-x}$(Mo$_{2}$C)$_{x}$ samples with $x$ up to 0.6.
The MoReRu ($x$ = 0) alloy consists of a single hexagonal close-packed phase (space group $P$6$_{3}$/$mmc$), in which all the Mo, Re and Ru atoms are distributed randomly at the single (1/3, 2/3, 0.25) site.\cite{lee2019superconductivity}
After the addition of Mo$_{2}$C, this hexagonal phase is retained up to $x$ = 0.3, at which a small peak near 2$\theta$ $\approx$ 26.6$^{\circ}$ due to carbon impurity is discernible.
The lattice parameters are determined by least-squares fittings.
For $x$ = 0, one obtains $a$ = 2.761 {\AA} and $c$ = 4.433 {\AA}, in line with those reported previously.\cite{lee2019superconductivity}
With the increase of $x$, the $a$- and $c$-axis increase monotonically to 2.796 {\AA} and 4.505 {\AA}, respectively.
Since the atomic radius of C (0.70 {\AA}) is considerably smaller than those of the transition metals Mo (1.40 {\AA}), Re (1.38 {\AA}), and Ru (1.35 {\AA}),\cite{atomicradius}
the observed lattice expansion implies that the C atoms most likely reside at the interstitial positions of the hexagonal lattice, similar to MoReRuC$_{x}$.\cite{zhu2022superconducting}
As $x$ is increased further, however, the hexagonal phase is destabilized.
The samples with $x$ = 0.4 and 0.6 apparently contain multiple phases with overlapped diffraction peaks.
Surprisingly, at the middle $x$ = 0.5, the diffraction peaks become sharp and are well indexed on the noncentrosymmetric cubic $\beta$-Mn type structure with the $P$4$_{1}$32 space group. Based on these results, a partial phase diagram for the MoReRu-Mo$_{2}$C system is constructed in Fig. 1b. One can see that, while both MoReRu and Mo$_{2}$C are centrosymmetric, a new noncentrosymmetric compound is found to form at the composition of Mo$_{3}$ReRuC ($x$ = 0.5). In fact, we have also prepared a few samples for $x$ just around 0.5, all of which turn out to be multiphased (see Figure S1 of the Supporting Information). Especially, the peak positions of the $\beta$-Mn type phase for $x$ = 0.45 are almost the same as those for $x$ = 0.5. It thus appears that Mo$_{3}$ReRuC has a very narrow homogeneity range.

\subsection{Structural refinement and chemical analysis of Mo$_{3}$ReRuC}
The structure of Mo$_{3}$ReRuC is refined based on the model of $\beta$-Mn type Mo$_{3}$Rh$_{2}$N,\cite{shang2018} where the Mo atoms occupy the 12$d$ (0.125, 0.206, 0.456) site, Rh the 8$c$ (0.061, 0.061, 0.061) site and C the 4$a$ (0.375, 0.375, 0.375) site.
In the case of Mo$_{3}$ReRuC, it is assumed that Re and Ru atoms are distributed randomly at the 8$c$ sites while the other atoms are the same as in Mo$_{3}$Rh$_{2}$N.\cite{shang2018}
The refinement profile displayed in Fig. 2a indicates a reasonably good agreement between the calculated and observed patterns, which is corroborated by the small $R_{\rm wp}$ (9.7\%) and $R_{\rm p}$ (7.3\%) factors as listed in Table 1.
Note that the possibility of site mixing between Mo and Ru atoms cannot be excluded considering their close atomic numbers.
In addition, the relatively large residual plot is probably due to the effects of crystallite size, lattice strain and surface roughness, which lead to change in the peak shape.
The refined cubic lattice parameter $a$ = 6.8107(1) {\AA} is nearly the same as that Mo$_{3}$Rh$_{2}$N\cite{shang2018} ($a$ = 6.81 {\AA}) but smaller than that of Mo$_{3}$Al$_{2}$C ($a$ = 6.87 {\AA})\cite{bauer2010unconventional,karki2010structure}.
This is as expected since the Re and Ru have atomic radii very similar to that of Rh while smaller than that of Al (1.42 {\AA}).\cite{atomicradius}
The schematic structure for Mo$_{3}$ReRuC and its project along the $c$-axis are plotted in Figs. 2b and c, respectively.
In particular, the Mo atoms form a three dimensional network of corner-connected metaprisms, whose centers are occupied by carbon atoms.

Figure 3a show the typical SEM image with a scale bar of 10 $\mu$m for the Mo$_{3}$ReRuC sample.
It is clear that the sample is dense and free from pores and voids.
Furthermore, the corresponding EDX elemental maps displayed in Figs. 3b-e reveal that all the Mo, Re, Ru, and C elements are distributed uniformly.
The average ratio of Mo:Re:Ru is 2.8(1):1.0(1):1.2(1), consistent with the nominal one.
Note that the EDX analysis cannot determine the carbon content accurately considering its small atomic number.
Nevertheless, given the low vapor pressure of carbon and its full occupancy from the structural refinement, the deviation in actual carbon content is expected to be negligible.
\begin{figure}
\includegraphics*[width=8cm]{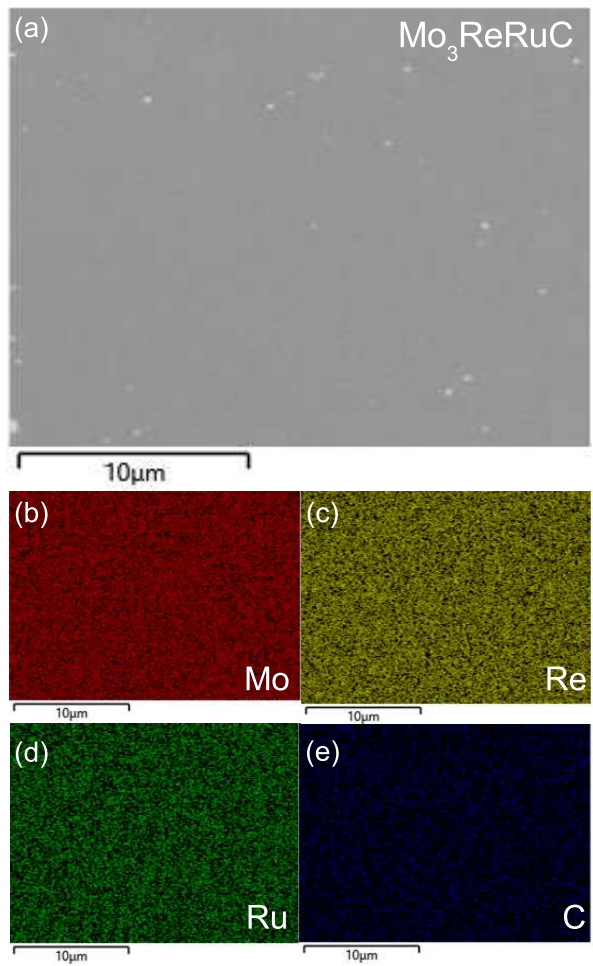}
\caption{
(a) Typical SEM image for Mo$_{3}$ReRuC on a scale bar of 10 $\mu$m.
(b-e) Corresponding EDX elemental mapping results for Mo, Re, Ru and C.
}
\label{fig3}
\end{figure}

\subsection{Superconductivity in Mo$_{3}$ReRuC}
The resistivity curve for the Mo$_{3}$ReRuC sample is shown in Fig. 4a and its inset.
With decreasing temperature, a weak metallic behavior is observed, in analogy with that observed in isostructural W$_{4}$IrC\cite{zhu2022w} and W$_{7}$Re$_{13}$C\cite{kawashima2005superconductivity}.
Below $\sim$7.8 K, the resistivity drops sharply to zero, indicating a superconducting transition.
Meanwhile, a clear diamagnetic transition is present in both ZFC and FC susceptibility curves, as shown in Fig. 4b.
Its coincides well with the midpoint of resistivity drop, and hence $T_{\rm c}$ is determined to be 7.7 K.
At 1.8 K, the shielding and Meissner fractions, 4$\pi$$\chi_{\rm ZFC}$ and 4$\pi$$\chi_{\rm FC}$, are $>$140\% and $<$30\%, respectively, without demagnetization correction.
This divergence indicates a type-II superconducting behavior.
\begin{figure}
\includegraphics*[width=8.5cm]{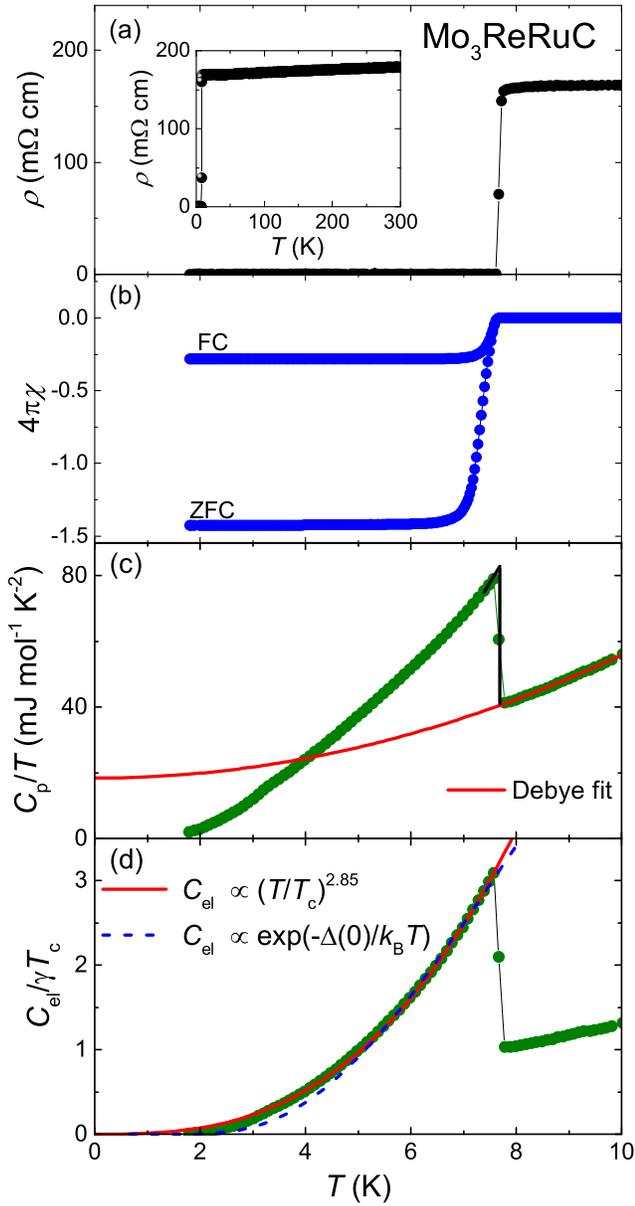}
\caption{
(a-c) Temperature dependencies of resistivity, magnetic susceptibility and specific heat below 10 K, respectively, for the Mo$_{3}$ReRuC sample.
In panel (a), the inset shows the resistivity data up to 300 K.
In panel (c), the red line a fit by the Debye model and the black ones are entropy conserving construction to estimate the size of specific heat jump.
(d) Temperature dependence of normalized electronic specific heat. Power-law and exponential fits to the data are denoted by the solid and dashed lines, respectively.
}
\label{fig4}
\end{figure}
As shown in Fig. 4c, the distinct specific heat ($C_{\rm p}$) anomaly confirms the bulk nature of superconductivity and its size is estimated to be $\Delta$$C_{\rm p}$/$T_{\rm c}$ = 42.2 mJ/molK$^{2}$ based on the entropy conserving construction.
Above $T_{\rm c}$, the normal-state $C_{\rm p}$ data follow well the Debye model,
\begin{equation}
C_{\rm p}/T = \gamma + \beta T^{2},
\end{equation}
where $\gamma$ is the Sommerfeld coefficient and $\beta$ is phonon specific-heat coefficient.
The best fit gives $\gamma$ = 18.36 mJ mol$^{-1}$ K$^{-2}$ and $\beta$ = 0.3742 mJ mol$^{-1}$ K$^{-4}$.
The normalized specific heat jump is then found to be $\Delta$$C_{\rm p}$/$\gamma$$T_{\rm c}$ $\approx$ 2.3, significantly larger than 1.43 expected from the BCS theoty \cite{bardeen1957theory}.
Also, the Debye temperature $\Theta_{\rm D}$ is calculated to be 315 K through the relation
\begin{equation}
\Theta_{\rm D} = (12\pi^{4} N R/5\beta)^{1/3},
\end{equation}
where $N$ = 6 is the number of atoms in the formula and $R$ = 8.314 J mol$^{-1}$ K$^{-1}$ is the molar gas constant.
The $\Theta_{\rm D}$ allows us to estimate the electron-phonon coupling strength $\lambda_{\rm ep}$ to be 0.73 based on the inverted McMillan formula\cite{mcmillan1968transition},
\begin{equation}
\lambda_{\rm ep}=\frac{1.04+\mu^{*} \ln \left(\Theta_{\rm D} / 1.45 T_{\mathrm{c}}\right)}{\left(1-0.62 \mu^{*}\right) \ln \left(\Theta_{\rm D} / 1.45 T_{\mathrm{c}}\right)-1.04},
\end{equation}
where $\mu^{*}$ = 0.13 is the Coulomb repulsion pseudopotential.
In passing, we have also measured physical properties of the (MoReRu)$_{1-x}$(Mo$_{2}$C)$_{x}$ sample with $x$ = 0.45 (see Figure S2 of the Supporting Information). Its resistive-transition onset temperature is very close to that of Mo$_{3}$ReRuC. However, the magnetic and specific heat data indicate that the (MoReRu)$_{0.55}$(Mo$_{2}$C)$_{0.45}$ sample contains multiple superconducting phases, in line with the multiple-phase nature found by XRD. These results further corroborate the stoichiometric nature of the Mo$_{3}$ReRuC sample.

\begin{figure}[h]
\includegraphics*[width=7.8cm]{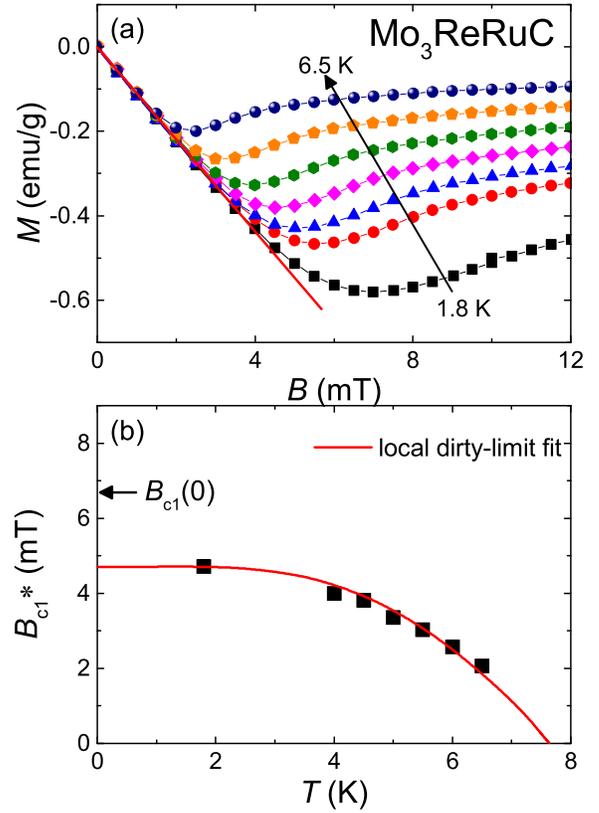}
\caption{
(a) Isothermal magnetization curves at temperatures between 1.8 and 6.5 K for the Mo$_{3}$ReRuC sample.
The initial linear behavior and the direction of temperature increase are marked by the solid line and arrow, respectively.
(b) Temperature dependence of the effective lower critical field. The solid line is a fit by the local dirty limit formula and the arrow marks the zero-temperature value after correction for the demagnetization effect.
}
\label{fig5}
\end{figure}
\begin{figure}[h]
\includegraphics*[width=8cm]{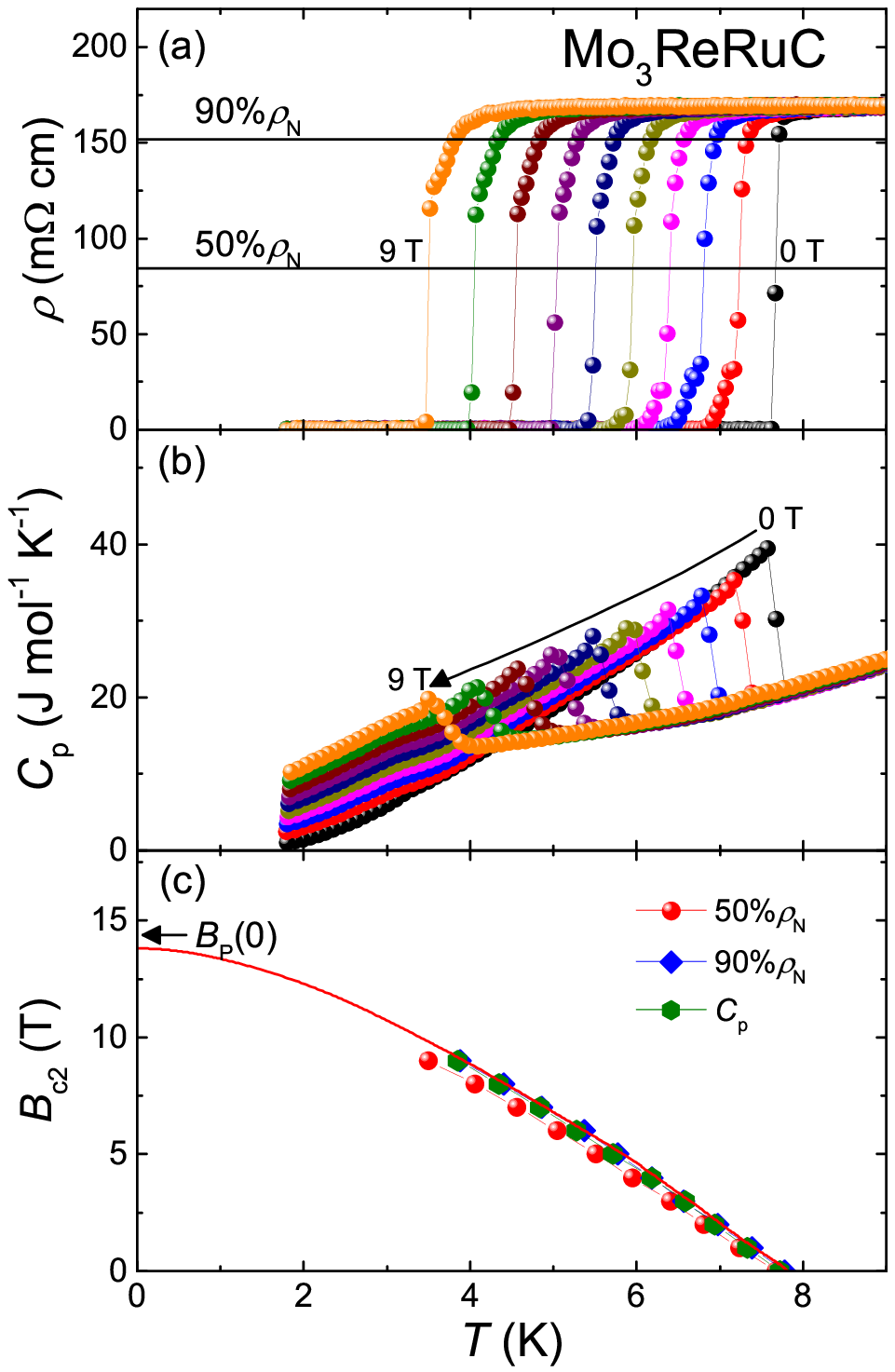}
\caption{
(a) Resistive transitions for Mo$_{3}$ReRuC under a magnetic field increment of 1 T up to 9 T.
The two horizontal lines correspond to 90\% and 50\% of the normal-state resistivity, respectively.
(b) Specific heat jumps under fields up to 9 T.
(c) Temperature dependencies of upper critical fields from both resistivity and specific heat measurements.
The solid line denotes a fit by the WHH model.
}
\label{fig6}
\end{figure}

The normalized electronic specific heat $C_{\rm el}$ is isolated by subtraction of the phonon contribution and plotted as $C_{\rm el}$/$\gamma$$T_{\rm c}$ in Fig. 4d.
It is found that the $C_{\rm el}$/$\gamma$$T_{\rm c}$ data match well the single-gap BCS-type model\cite{bardeen1957theory} $C_{\rm el}$/$\gamma$$T_{\rm c}$ $\propto$ exp(-$\Delta$(0)/$k_{\rm B}T$) in the temperature ranges below $\sim$2.5 K and between $\sim$5.3 K and $T_{\rm c}$.
In the intermediate temperatures, the $C_{\rm el}$/$\gamma$$T_{\rm c}$ data grow faster than the exponential function and appear to follow a power-law behavior $C_{\rm el}$/$\gamma$$T_{\rm c}$ $\propto$ $T^{2.85}$, which actually extends up to $T_{\rm c}$.
This is reminiscent of those observed in Mo$_{3}$Al$_{2}$C\cite{bauer2010unconventional} and W$_{3}$Al$_{2}$C\cite{ying2019superconductivity}, where it is attributed to the presence of gapless excitation.
Nevertheless, it is pointed out that the deviation from the exponential behavior is not incompatible with the single-gap BCS behavior since the fit often fails in the temperature range where the gap magnitude shows a sizable variation.
Hence, to draw a more definitive conclusion on the gap structure, further experiments such as penetration depth and tunneling spectra measurements are called for.

To determine the lower critical field $B_{\rm c1}$ of Mo$_{3}$ReRuC, isothermal magnetization curves are measured between 1.8 K and 6 K and plotted in Fig. 5a.
At each temperature, the magnetization decreases linearly with initially increasing magnetic field, consistent with a Meissner state.
The effective lower critical field $B_{\rm c1}^{\ast}$ is defined as the field where magnetization curve starts to deviate this linearity.
The zero-temperature effective lower critical field $B_{\rm c1}^{\ast}$(0) is obtained to be 4.7 mT by extrapolating the data using the local dirty limit formula.
The sample used for the magnetic measurements has an irregular shape, which makes it difficult to estimate the demagnetization factor $N_{\rm d}$ directly.
Nevertheless, from the initial slope of magnetization curve, $N_{\rm d}$ can be derived as
\begin{equation}
\rm d \it M/\rm d \it B = \rm -\frac{1}{4\pi(1 - \it N_{\rm d})},
\end{equation}
which gives $N_{\rm d}$ = 0.3. Since Then the zero-temperature lower critical field $B_{\rm c1}$(0) = $B_{\rm c1}^{\ast}$(0)/(1 $-$ $N_{\rm d}$), one obtains $B_{\rm c1}$(0) = 6.7 mT for Mo$_{3}$ReRuC.

The upper critical field $B_{\rm c2}$ of Mo$_{3}$ReRuC is obtained by resistivity and specific heat measurements under magnetic fields up to 9 T, and the results are shown in Fig. 6a and b.
In both cases, the field increment is 1 T and the gradual suppression of superconducting transition is evident.
Especially, the resistive transition is broadened under field and a resistivity tail is observed in the filed interval between 1 T and 4 T.
The latter is commonly observed for $\beta$-Mn type superconductors and attributed to the poor conducting grain boundaries\cite{zhu2022w}.
Hence, for the resistivity measurements, we define two transition temperatures at each field, $T_{\rm c}^{50\%\rho_{\rm N}}$ and $T_{\rm c}^{90\%\rho_{\rm N}}$, which correspond to 50\% and 90\% of the normal-state resistivity, respectively.
Also, $T_{\rm c}^{C_{\rm p}}$ is defined as the onset of $C_{\rm p}$ jump.
The resulting $B_{\rm c2}$ versus $T$ diagrams are summarized in Fig. 6c.
Here the correction of $B_{\rm c2}$ due to the demagnetization effect is negligible since the shield fraction at 1.8 K is only $\sim$0.1\% at 1 T and further decreased to $\sim$0.003\% at 5 T.
One can see that $B_{\rm c2}^{90\%\rho_{\rm N}}$($T$) agrees well with $B_{\rm c2}^{C_{\rm p}}$($T$) and both are larger than $B_{\rm c2}^{50\%\rho_{\rm N}}$($T$).
Since specific heat is not sensitive to the grain boundary effects, the former two sets of data should be more reliable.
Extrapolating these data to 0 K using the Werthamer-Hohenberg-Helfand model\cite{werthamer1966temperature} yields the zero-temperature upper critical field $B_{\rm c2}$(0) = 13.8 T and a $B_{\rm c2}$(0)/$B_{\rm P}$(0) ratio of 0.97, where $B_{\rm P}$(0) = 1.86$T_{\rm c}$ $\approx$ 14.2 T is the Pauli paramagnetic limit \cite{clogston1962upper}. This ratio is the same as that of W$_{4}$IrC$_{0.8}$ \cite{zhu2022w} and among the highest for $\beta$-Mn type NCSs.

We now evaluate various superconducting parameters with the values of $B_{\rm c1}$(0) and $B_{\rm c2}$(0).
The Ginzburg-Landau (GL) coherence length $\xi_{\rm GL}$ is calculated to be 4.9 nm according to the equation
\begin{equation}
\xi_{\rm GL}(0) = \sqrt{\frac{\Phi_{0}}{2\pi B_{\rm c2}(0)}},
\end{equation}
where $\Phi_{0}$ = 2.07 $\times$ 10$^{-15}$ Wb is the flux quantum.
Given that
\begin{equation}
\frac{B_{\rm c1}(0)}{B_{\rm c2}(0)} = \frac{\rm ln\kappa_{\rm GL}+0.5}{2\kappa_{\rm GL}^{2}},
\end{equation}
we obtain the GL parameter $\kappa_{\rm GL}$ = 70.7, affirming that Mo$_{3}$ReRuC is a type-II superconductor.
With $B_{\rm c1}$(0) and $\kappa_{\rm GL}$, the penetration depth $\lambda_{\rm eff}$ is found to be 346 nm using the formula:
 \begin{equation}
B_{\rm c1}(0) = \frac{\Phi_{0}}{4\pi\lambda_{\rm eff}^{2}}(\rm ln\kappa_{\rm GL}+0.5).
\end{equation}
\begin{table*}
	\centering
	\caption{Physical parameters for Mo$_{3}$ReRuC, together with the data for Mo$_{3}$Al$_{2}$C and W$_{3}$Al$_{2}$C for comparison.}
	\renewcommand\arraystretch{1.3}
	\begin{tabular}{p{2cm}<{\centering}p{3cm}<{\centering}p{2cm}<{\centering}p{2cm}<{\centering}p{2cm}<{\centering}}
		\\
        \hline
         Parameter &Unit& Mo$_{3}$ReRuC & Mo$_{3}$Al$_{2}$C & W$_{3}$Al$_{2}$C\\
        \hline
        $T_{\rm c}$ & K&  7.7&  9.2 &  7.6\\
        $\gamma$  & mJ mol$^{-1}$ K$^{-2}$&18.36 &18.65 &7.3\\
        $\Delta$$C_{\rm p}$/$\gamma$$T_{\rm c}$  & $-$&2.3 &2.14 &2.7\\
        $\delta$ &mJ mol$^{-1}$ K$^{-4}$& 0.3742 & 0.3050 & 0.62\\
        $\Theta_{\rm D}$ & K &315 &338.5 &266\\
        $\lambda_{\rm ep}$  & $-$&0.73&$-$ &$-$\\
        $B_{\rm c1}$ & mT &6.7&4.7&10\\
        $B_{\rm c2}$ & T &13.8&15.1-15.6&$>$10\\
        $\lambda_{\rm eff}$ & nm &346&375.5&$<$320\\
        $\xi_{\rm GL}$ & nm &4.9&4.2&$<$5.7\\
        $\kappa_{\rm GL}$ & $-$ &70.7&88.6&56\\
		\hline
	\end{tabular}
	\label{Table2}
\end{table*}
\subsection{Theoretical band structure of Mo$_{3}$ReRuC}
\begin{figure}
\includegraphics*[width=9cm]{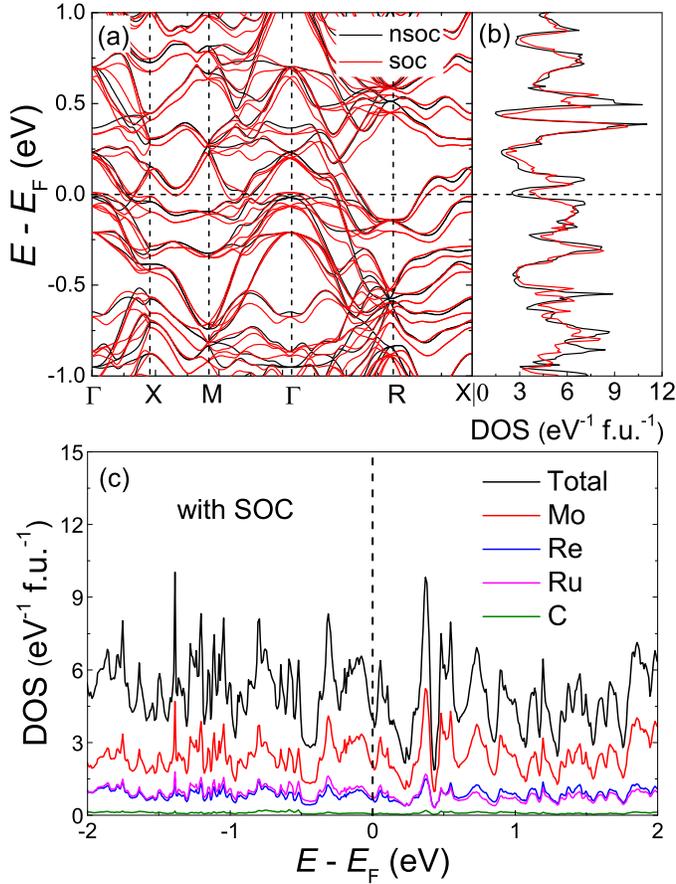}
\caption{
(a) Calculated band structure of Mo$_{3}$ReRuC without (black) and with (red) spin-orbit coupling (SOC).
(b) Corresponding energy dependence of the total density of states.
In both panels, the position of the Fermi level is indicated by the horizontal dashed line.
(c) Projection of density of states with SOC on different atoms. The position of the Fermi level is marked by the vertical line.
}
\label{fig6}
\end{figure}
Figures 7a and b show the theoretical band structure and corresponding energy dependencies of DOS for Mo$_{3}$ReRuC.
In the calculation, we employed a hypothetical crystal structure of Mo$_{3}$ReRuC, in which the total eight 8$c$ sites are occupied orderly by four Re atoms and four Ru atoms. This configuration lowers the space group symmetry from $P$4$_{1}$32 to P2$_{1}$3.
Without spin-orbit coupling (SOC), multiple bands are crossing the Fermi level ($E_{\rm F}$), which lies close to a local DOS minimum.
When turning on SOC, the number of bands doubles due to the lift of degeneracy and the band dispersion near $E_{\rm F}$ changes significantly.
As a consequence, the $E_{\rm F}$ is pushed slightly away from the local minimum and the DOS at $E_{\rm F}$ is increased by $\sim$ 60\% from 2.52 eV$^{-1}$ per f.u. to 4.03 eV$^{-1}$ per f.u.
The latter corresponds to a bare electronic specific heat coefficient $\gamma^{\rm bare}$ = 9.3 mJ mol$^{-1}$ K$^{-2}$.
Then the $\lambda_{\rm ep}$ can be given as $\lambda_{\rm ep}$ = $\gamma$/$\gamma^{\rm bare}$ $-$ 1 = 0.97.
Since the electronic correlation can also contribute to the enhancement in $\gamma$,
this value tends to be overestimated and hence should be taken as consistent with that deduced from the C$_{p}$ measurement.
The atomic projected DOS with SOC for Mo$_{3}$ReRuC is shown in Fig. 7c.
In the $E$ $-$ $E_{\rm F}$ range from $-$2 to 2 eV, the DOS is dominated by the Mo states, followed by almost equal Re and Ru states.
At $E_{\rm F}$, the Mo contribution is nearly triple those of Re and Ru atoms.
In contrast, the contribution from C atoms is negligible over the whole $E$ $-$ $E_{\rm F}$ range investigated.
This is consistent with the common belief that the light elements such as carbon mainly play a structural role in stabilizing the $\beta$-Mn type structure.

\subsection{Comparison with other $\beta$-Mn type NCSs}
The overall band structure of Mo$_{3}$ReRuC to a large extent resembles those of Mo$_{3}$Al$_{2}$C\cite{bauer2010unconventional} and W$_{3}$Al$_{2}$C\cite{ying2019superconductivity}, except that the contribution from Al atoms is negligible in the latter two cases.
Hence, to gain insight in the factors determining $T_{\rm c}$, we present a comparison between their physical parameters as listed in Table 2.
One can see that the $T_{c}$ of Mo$_{3}$ReRuC is lower than that of Mo$_{3}$Al$_{2}$C\cite{bauer2010unconventional} (9.2 K) but very close to W$_{3}$Al$_{2}$C (7.6 K)\cite{ying2019superconductivity} .
Nonetheless, the $\gamma$ values of Mo$_{3}$ReRuC (18.36 mJ mol$^{-1}$ K$^{-2}$) and Mo$_{3}$Al$_{2}$C\cite{bauer2010unconventional} (18.65 mJ mol$^{-1}$ K$^{-2}$) are more than twice that of W$_{3}$Al$_{2}$C\cite{ying2019superconductivity} (7.3 mJ mol$^{-1}$ K$^{-2}$). According to McMillan\cite{mcmillan1968transition}, the $\lambda_{\rm ep}$ can be expressed as
\begin{equation}
\lambda_{\rm ep} \equiv \frac{N(0)\langle I^{2}\rangle}{M\langle \omega^{2}\rangle},
\end{equation}
where $N$(0) is the bare density of states at $E_{\rm F}$, $M$ is the atomic mass, $\langle$$I^{2}$$\rangle$ is the averaged electron-phonon matrix element and $\langle$$\omega^{2}$$\rangle$ is the averaged phonon frequency.
Assuming that $\langle$$I^{2}$$\rangle$ and $\langle$$\omega^{2}$$\rangle$ are the same for the three compound, one would expect that the $\lambda_{\rm ep}$ values of Mo$_{3}$ReRuC and Mo$_{3}$Al$_{2}$C are of similar magnitude and about twice that of W$_{3}$Al$_{2}$C.
Consequently, $T_{\rm c}$ of the former two compounds would be considerably higher than that of the latter, which is clearly at odds with the experimental observations.
Hence it is more plausible that the DOS at $E_{\rm F}$ does not play a dominant role in determining $\lambda_{\rm ep}$ and hence $T_{\rm c}$ in these compounds.

\section{Conclusions}
To conclude, we have discovered a quaternary compound Mo$_{3}$ReRuC when exploring the MoReRu-Mo$_{2}$C system.
Mo$_{3}$ReRuC adopts a noncentrosymmetric cubic $\beta$-Mn type structure and exhibit bulk type-II superconductivity below 7.7 K.
In particular, the ratio of $B_{\rm c2}$(0)/$B_{\rm P}$(0) approaches one and the specific heat jump implies a strongly coupled superconducting state.
Band structure calculations show that the Fermi level of Mo$_{3}$ReRuC lies close to a local minimum of the DOS, which is mainly contributed from the Mo, Re and Ru atoms and strongly increased by the spin-orbit coupling.
Our results call for further investigations of the intermediate phase between alloys and carbides, which may offer an alternative strategy to look for new $\beta$-Mn type NCSs.

\section*{ACKNOWLEDGEMENT}
We acknowledge the foundation of Westlake University for financial support.
The work at Zhejiang University is supported by the National Natural Science Foundation of China (No.12050003).

\end{document}